\documentstyle[11pt,emulateapj5]{aastex}
 
\newcommand{\ngal}{334}

\newcommand{\CII}{\hbox{{\rm C}\kern 0.1em{\sc ii}}}
\newcommand{\CIV}{\hbox{{\rm C}\kern 0.1em{\sc iv}}}

\newcommand{\FeI}{\hbox{{\rm Fe}\kern 0.1em{\sc i}}}
\newcommand{\FeII}{\hbox{{\rm Fe}\kern 0.1em{\sc ii}}}
\newcommand{\SiII}{\hbox{{\rm Si}\kern 0.1em{\sc ii}}}
\newcommand{\AlII}{\hbox{{\rm Al}\kern 0.1em{\sc ii}}}
\newcommand{\NiII}{\hbox{{\rm Ni}\kern 0.1em{\sc ii}}}
\newcommand{\CrII}{\hbox{{\rm Cr}\kern 0.1em{\sc ii}}}

\newcommand{\ZnII}{\hbox{{\rm Zn}\kern 0.1em{\sc ii}}}
\newcommand{\OI}{\hbox{{\rm O}\kern 0.1em{\sc i}}}
\newcommand{\MgI}{\hbox{{\rm Mg}\kern 0.1em{\sc i}}}
\newcommand{\MgII}{\hbox{{\rm Mg}\kern 0.1em{\sc ii}}}

\newcommand{\HI}{\hbox{{\rm H}\kern 0.1em{\sc i}}}
\newcommand{\HII}{\hbox{{\rm H}\kern 0.1em{\sc ii}}}
\newcommand{\lya}{\hbox{{\rm Ly}\kern 0.1em$\alpha$}}
\newcommand{\Ly}{\hbox{{\rm Ly}\kern 0.1em$\alpha$}}
\newcommand{\Ha}{\hbox{{\rm H}\kern 0.1em$\alpha$}}
\newcommand{\Hb}{\hbox{{\rm H}\kern 0.1em$\beta$}}

\newcommand{\mpy}{\hbox{$M_{\odot}$~yr$^{-1}$}}

\newcommand{\msun}{\hbox{$M_{\odot}$}}

\newcommand{\Ldens}{\hbox{ergs s$^{-1}$ Hz$^{-1}$}}

\newcommand{\hMpc}{\hbox{$h^{-1}$~Mpc}}

\newcommand{\kms}{km~s$^{-1}$}

\slugcomment{Received 2004 November 7; accepted 2005 March 8; published ApJL 2005 April 20}

\begin{document}

\title{The Star Formation Rate-Density Relationship at Redshift 3}

\author{Nicolas Bouch\'e~\altaffilmark{1}}
\affil{European Southern Observatory, Karl-Schwarzschild-Strasse 2, D-85748 Garching, Germany;
and Max Planck Institut f\" ur  extraterrestrische Physik, Giessenbachstrasse, D-85748 Garching, Germany;
nbouche@mpe.mpg.de}
\author{James D. Lowenthal~\altaffilmark{1}}
\affil{Five College Astronomy Department, Smith College, McConnell Hall, Northampton, MA 01063;
james@ast.smith.edu}

\altaffiltext{1}{Visiting Astronomer, Kitt Peak National Observatory, National Optical
    Astronomy Observatory, which is operated by the Association of
    Universities for Research in Astronomy, Inc., under cooperative
    agreement with the National Science Foundation.}
 
\begin{abstract}
We study  the star formation rate (SFR) as a function of 
environment for UV-selected Lyman break galaxies (LBGs) at redshift 3.
From   deep  ($\mu_{I,\rm AB}(sky)\simeq 27.6$)   $UBVI$ KPNO 4-m/MOSAIC images,
  covering a total of 0.90 deg$^2$,  
  we select  \ngal\ LBGs in  slices  $100$\hMpc\ (co-moving) deep spanning
  the redshift range  $2.9<z<3.4$ based on
 Bayesian photometric redshifts  that include the   $I$ magnitude as a prior.
 The slice width ($100$\hMpc) corresponds to the photometric redshift accuracy ($\Delta_z\sim 0.15$).
 We used mock catalogs from the GIF2 cosmological simulations to show that this redshift resolution  is sufficient to
 statistically differentiate the high-density regions from the low-density
 regions using  $\Sigma_5$,  the projected  density to the fifth nearest neighbor.
These mock catalogs have a redshift depth  of $110$\hMpc, similar to our slice width.
The large area of the MOSAIC images,   $\sim40\times 40$~Mpc (co-moving) per field,
allows us to measure   the SFR from the dust-corrected UV continuum    
as a function of $\Sigma_5$.  In contrast to low-redshift galaxies,
 we find that   the SFR (or UV luminosity) of LBGs at $z = 3$ shows no detectable
dependence on environment over   2 orders of magnitude in density.
To test the significance of our result, we use
 Monte Carlo simulations (from the mock catalogs)  and the same projected density estimators 
 that we applied to our data.
 We find that we can reject  the steep $z=0$ SFR-density relation at the 5$-\sigma$ level.
 We conclude  that  the SFR-density relation at $z=3$ must be at least 3.6 times flatter than it is locally,
 i.e. the SFR of LBGs   is significantly less dependent on environment than
 the SFR of local star-forming galaxies.
We find that the rest-frame UV colors are also independent of environment.
\end{abstract}



\section{Introduction}

There is a long history of results showing that galaxy properties vary as a function of
environments such as the morphology-density relation  in clusters
\citep{MelnickJ_77a,DresslerA_80a,PostmanM_84a,DresslerA_97a,TreuT_03a}
and its associated star formation rate-density relation \citep[e.g.,][]{LewisI_02a}.
\citet{LewisI_02a} used the Two-Degree Field Galaxy Redshift Survey \citep[2dFGRS;][]{CollessM_01a} to show that
the anti-correlation between star formation rate (SFR) and local projected density holds
in clusters up to 2 virial radii from the cluster center.
Not only is the mean luminosity  a function of environment
\citep{BlantonM_03b,HoggD_03a,BaloghM_04b}, but the luminosity function (LF)
is also a continuous function of the local density \citep[e.g.][]{CrotonD_04a}. 
  
The SFR also depends on the local density   
in groups and in the field.
\citet{GomezP_03a} showed similar results  using the Early Data Release \citep{StoughtonC_02a}
of  the Sloan Digital Sky Survey (SDSS), and they also showed that the SFR-density relation holds for field galaxies.
\citet{BaloghM_04a} used both the 2dFGRS and the SDSS to show that 
the SFR (as measured by the \Ha\ equivalent width) of field galaxies is strongly dependent
on the local projected density.
All these results point to the existence of physical mechanisms that
 quench  star formation as the local density increases, from   the
 field to groups to clusters.

At  $z=0.4$, the quenching of star-formation is apparently already in place.
 Using the  wide-field  ($\sim 30$\arcmin) imager SuprimeCam at the Subaru 8-m
 telescope,  \citet{KodamaT_01a} selected    members of
  the $z=0.41$ cluster A851   using photometric redshifts and
 found that the colors of faint galaxies are strongly dependent on
the  local galaxy density. Follow-up narrow band \Ha\ imaging of this cluster
 showed that  the  fraction of star-forming galaxies is a strong function of the local
projected density  far ($\ga$1\hMpc) from the cluster center \citep{KodamaT_04a}. 

No   observations of the SFR-density relation have yet been made at redshifts  greater than one.
Here, we used ground-based data from the Kitt Peak 4--m telescope fitted with the MOSAIC
wide-field imager to investigate whether or not
the SFR-density relation already existed at redshift $z=3$.
We use photometric redshifts that are accurate to $\delta_z \sim 0.15$
or $\sim 100~h^{-1}$~Mpc, comparable in co-moving distance to those of the COMBO-17 survey.

We present our data and briefly discuss our selection of Lyman break galaxies (LBGs) in section~\ref{section:data}.
We then use mock catalogs from the GIF2 collaboration \citep{GaoL_04a}  
to show in section~\ref{section:simulations} that our photometric redshift accuracy  is sufficient
to statistically distinguish between low and high density regions. 
We estimated the SFR as in section~\ref{section:SFR}.
Finally, our results are presented in section~\ref{section:results}
and conclusions in section~\ref{section:discussion}.

Throughout this Letter, we adopt $\Omega_M=0.3$, $\Omega_\Lambda=0.7$
and $H_0=100~h$~\kms~Mpc$^{-1}$.
 At redshift $z\sim3$,
$H(z)\sim 4.46 H_0$, so $\delta z=0.1$ corresponds to 
$67~h^{-1}$~Mpc in co-moving coordinates.

\section{Data}
\label{section:data}

The observations used here are described in  detail in \citet{BoucheN_03b} 
and \citet[][hereafter BL04]{BoucheN_04c}. Briefly, we used
 the wide-field MOSAIC camera \citep{JacobyG_98a} at the Kitt Peak
National Observatory  4m telescope to image
 three fields in $UBVI$ to a limiting magnitude  $\mu_{I,\rm AB}(sky)\simeq 27.6$.
 These fields were  selected for the presence
of a damped \lya\ absorber (DLA) at $z\sim 3$ in our original DLA survey \citep[][BL04]{BoucheN_03a}.
Given the camera's field of view of 36\arcmin$\times$36\arcmin  (or $\sim40~h^{-1}$~Mpc on a side),
 our survey covers  a total of 0.90 deg$^2$.
We selected   $\sim$3,000 LBGs within the redshift range  $2.8<z_{\rm phot}<3.5$.
From Monte Carlo simulations,
our 50\% completeness level is  $I_{\rm AB}=24.8$ (${\mathcal R}\simeq 25$),
 corresponding    to $\sim 0.7 L^*$, using $L^*(z=3)$ from \citet{SteidelC_99a}.

 Photometric redshifts were obtained from the algorithm
 Hyperz \citep{BolzonellaM_00a} coupled with the prior likelihood distribution 
  given by the magnitude $m_I$ of each galaxy  \citep[as in][]{BenitezN_00a}.
See \citet{BoucheN_03b} and BL04 for a detailed description
of  the tests we performed on the algorithm  using spectroscopic redshifts in the Hubble Deep Field-North. 
 Up to $z<6$, the overall  rms  $\Delta_z$ is $0.1$--$0.15$, corresponding  to $\sim 67$--$100$~\hMpc,
 and  $\Delta_z/(1+z_{\rm spec})$ is $0.06$.

In order to selected    LBGs in a narrow redshift slice, we used the criterion:
\begin{equation}
 P(z_{\rm 0}\pm W_z/2) >0.5,\label{eq:slicecut}
\end{equation}
where $z_0$ is the slice redshift, $W_z$ is the redshift slice width, and $P(z)$
is the redshift probability distribution (BL04). 
  We choose a redshift width of $W_z=0.15$ because, as discussed in \citet{BoucheN_03a}, it produces
 the largest sample in the smallest redshift slice, given the rms of the photo-$z$'s.
At $z_{\rm 0}=3$, there are 68, 162, and 104 galaxies in our three fields for a total of \ngal.

\section{Can we use photometric redshifts to measure $\Sigma_5$?}
\label{section:simulations}

 Before investigating the galaxy properties as a function of environment, we
used the GIF2 numerical simulations of galaxy formation
(described in \citeauthor{GaoL_04a}~\citeyear{GaoL_04a} and
\citeauthor*{DeLuciaG_04a}~\citeyear{DeLuciaG_04a}) to investigate
whether or not, with such a coarse redshift resolution,
 one can make reliable density estimations based on  projected surface densities.
These $N$-body simulations were run with 400$^3$ particles
(each $1.73\times10^9$\msun)   over a periodic cube 
of 110~\hMpc\ on a side in a $\Lambda$ dominated universe.
From the output at $z=3$
containing $\sim 29,000$ galaxies, we 
 simulated an LBG sample by selecting 975 objects with
$m_{\rm I}< 25$ and 
$(B-V)< 0.2 $, 
where   $m_{\rm I}$ is the observed $I$ magnitude given the absolute magnitude
$M_I$, the redshift of the simulation $z=3$, and the $K$-correction according to
the type of spectral energy distribution of the galaxy.

\begin{figure}
\plotone{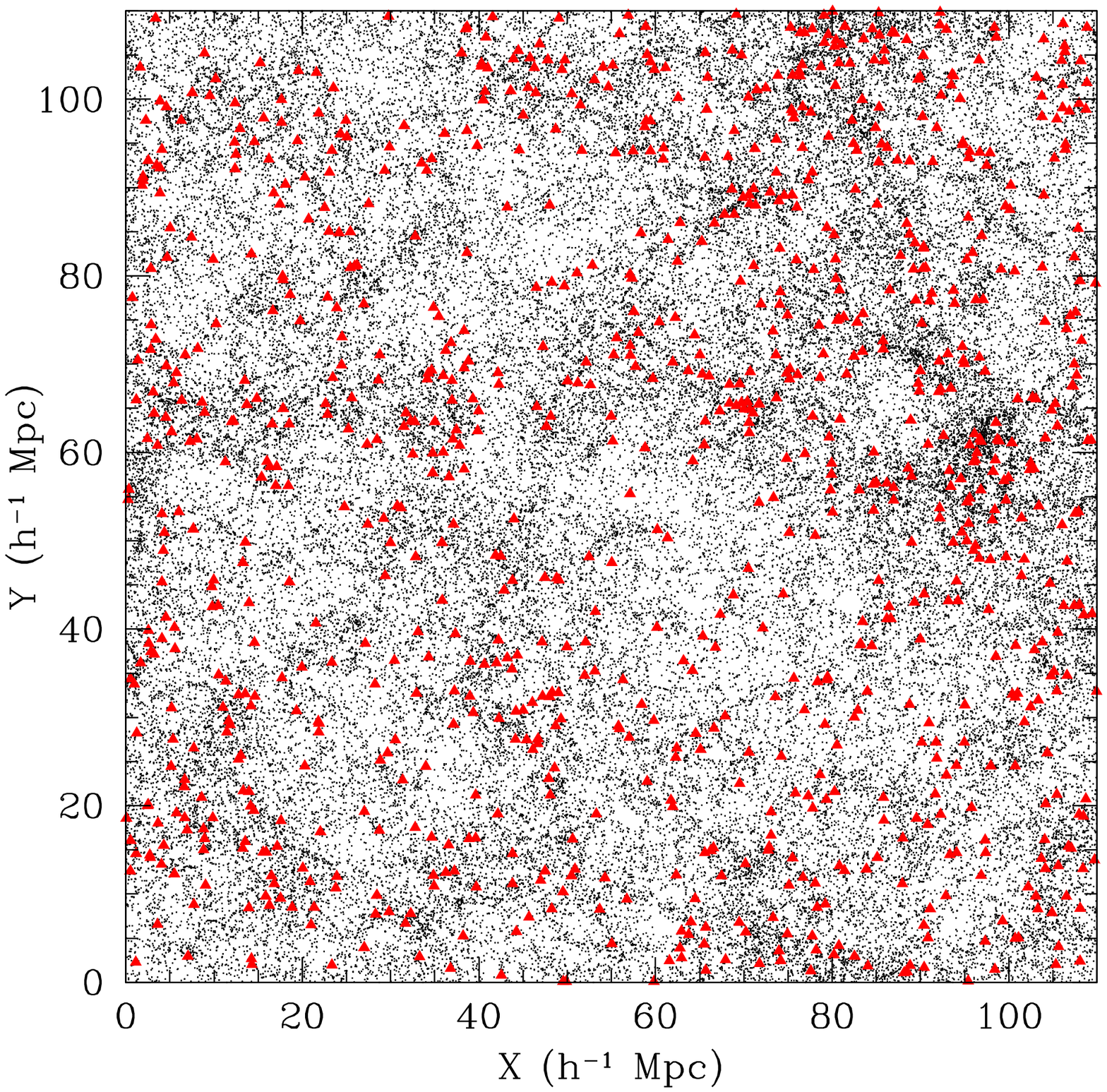}
\caption{The $X$--$Y$ distribution (dots) of 50\%\ (randomly selected) of the simulated galaxies of
the GIF2 simulations \citep{GaoL_04a}. The filled triangles show all the simulated LBGs.
Features of large-scale structure are visible in the
  simulated LBG population, considering that the galaxy positions are   projected along the $z$-direction over
 $\sim 100h^{-1}$~Mpc. This is quantified in Fig~\ref{fig:compare:n5:n10}.
\label{fig:2Dmap} }
\end{figure}

Fig.~\ref{fig:2Dmap} shows
the two-dimensional distribution (projected along the $z$-axis)
of simulated LBGs (filled triangles) along with all the galaxies (points)
in the simulated GIF2 volume. 
One sees that the simulated LBGs do trace the major over-dense regions
and filaments. Importantly, this `slice' width corresponds
to our redshift resolution, i.e. $\sim 100$~\hMpc.

Fig.~\ref{fig:compare:n5:n10}  shows, for the GIF2 simulation,
 the two-dimensional surface density, $\Sigma_5$,
 versus the three-dimensional true density $\rho_5$.
Both $\Sigma_5$ and $\rho_5$ are 
computed from the distance to the fifth nearest neighbor.
The horizontal error bars reflect the $1\sigma$ rms to the mean.
One order of magnitude change in $\Sigma_n$ corresponds to about
$1.5$ orders of magnitude in $\rho_n$. Thus,  we conclude
that the projected density 
gives a statistical measure of the true  three-dimensional over-density
 even when it is projected over 110~\hMpc.

\section{Estimating Star Formation Rates}
\label{section:SFR}

As reviewed by \citet{KennicuttR_98a},  one can estimate
the  SFR from the rest--frame UV luminosity density $L_\nu$ in the range $1500$--$2500$~\AA:
\begin{equation}
{\rm SFR}(\mpy)=1.4\times10^{-28} L_{\nu}(\Ldens) \label{eq:kennicutt}
\end{equation}
for a Salpeter initial mass function, covering the range 0.1--100~\msun.
This relation  applies only to galaxies with continuous star formation over time scales of $10^8$ yr or longer.
\citet{PettiniM_01a} have shown that for a small sample of LBGs, estimates of SFR
from Eq.~\ref{eq:kennicutt} are consistent with the SFR inferred   from \Hb\ line fluxes.

We estimated the  rest--frame luminosity densities (in \Ldens) of the \ngal\ LBGs in our MOSAIC images
from  the standard expression
$f_\nu=10^{0.4\cdot(m_{\rm AB}+48.6)}/(1+z_{\rm phot})$ and  
from the luminosity distance. 
The luminosity was corrected for dust extinction. 
The amount of dust extinction ($A_V$) is estimated simultaneously
 with the estimation of the photometric redshift  $z_{\rm phot}$.
 However, we found, using mock catalogs, that $A_V$ recovered by Hyperz is biased:
 $A_V(\hbox{true})=0.51\;A_V+0.32\label{eq:correction}\,.$
These mock catalogs were created from the same galaxy templates, with the same extinction
law used in  the redshift fitting and restricted to our redshift range of interest $2.85<z<3.25$.

We used the  $I$-band  ($\lambda_{\rm c}=8214$~\AA) flux to estimate the SFR; this filter
 corresponds to a rest-frame wavelength of 2000~\AA\ for $z=3$ galaxies.
We used the \citet{CalzettiD_00a} extinction curve~\footnote{This is the same extinction law
that was used by Hyperz in determining the photometric redshifts.}
to estimate $A_{2000\AA}$, and the dust corrected SFR ($\rm SFR_0$) follows from 
$L_{\rm \nu,o}=L_{\rm \nu}  10^{0.4\cdot A_{\nu}} $
and  Eq.~\ref{eq:kennicutt}.

\section{Results and Discussions}
\label{section:results}

For galaxies at  $z_{\rm phot}=3\pm 0.075$ selected with Eq.~\ref{eq:slicecut},
Fig.~\ref{fig:SFR-density} shows the dust-corrected star formation rate (SFR$_0$) as a function of the
normalized local density $\Sigma_5/<\Sigma_5>$ where  $\Sigma_5$  is the surface density within the
5th nearest neighbor, i.e. $\Sigma_5=6/(\pi r_5^2)$, and $<\Sigma_5>$ is its median
in each of our three MOSAIC fields.
The local  over-density $\delta_5$ is $\Sigma_5/<\Sigma_5>-1$.  
We use the normalized density    in order to combine
our several fields given that they  do not contain the same mean number of LBGs per unit area.
Error bars are shown for  20 randomly selected galaxies.
There is no detectable difference between
  the distribution of SFRs of LBGs in low-density environments and
  those in high-density environments.

We used the Kolmogorov-Smirnov (KS) test to determine whether or not the galaxies
in high density regions with $\delta_5>x$ are
a random subset of those at  $\delta_5<0$, i.e. they have the same SFR distribution.
The top panel of Fig.~\ref{fig:SFR-density} shows that we cannot reject the null hypothesis:
the sample with $\delta_5<0$ and the sample with $\delta_5>0$ are indistinguishable.
Similarly, a Pearson correlation test gives a $P$-value of $0.67$.
 In other words, there could be a correlation only at the 32\%\ confidence level.
 We repeated the analysis for higher redshift slices, ($z_{\rm 0}=3.1$, 3.2, and 3.3) and find similar results.

\begin{figure}
\plotone{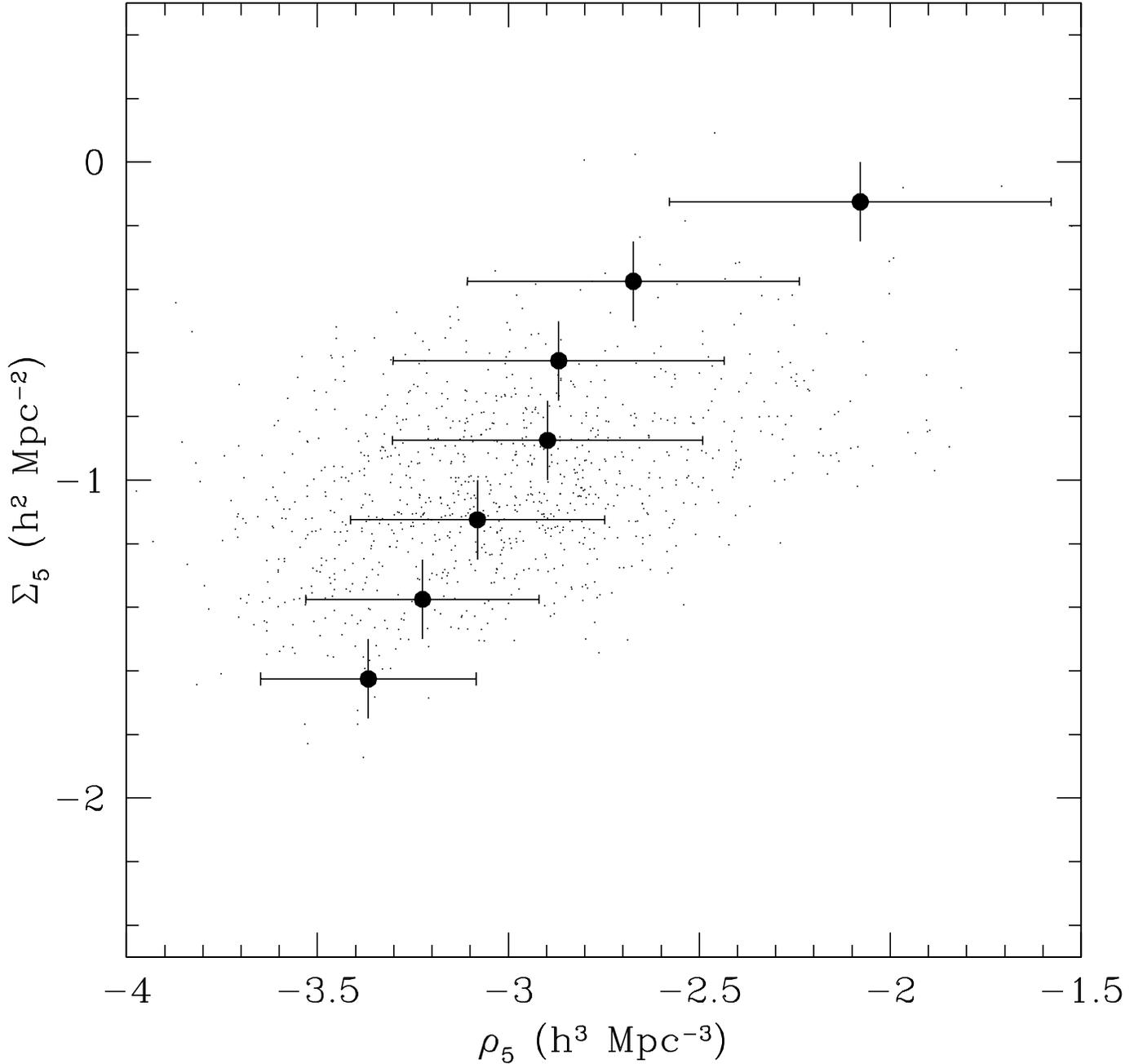}
\caption{For simulated LBGs in the GIF2 simulation, 
the projected two-dimensional density $\Sigma_5$ is plotted as a
function of  the three-dimensional density $\rho_5$. Both $\Sigma_5$ and $\rho_5$ are computed
from the fifth nearest neighbor.
 The points with error bars show the mean $\rho_5$ and its spread (rms) in bins
of $\Sigma_5$.
This shows that, even over 110~$h^{-1}$~Mpc, 
one can discriminate {\it statistically} between low- and high-density 
regions using projected density estimators such as 
$\Sigma_5$.
\label{fig:compare:n5:n10} }
\end{figure}

\begin{figure}
\plotone{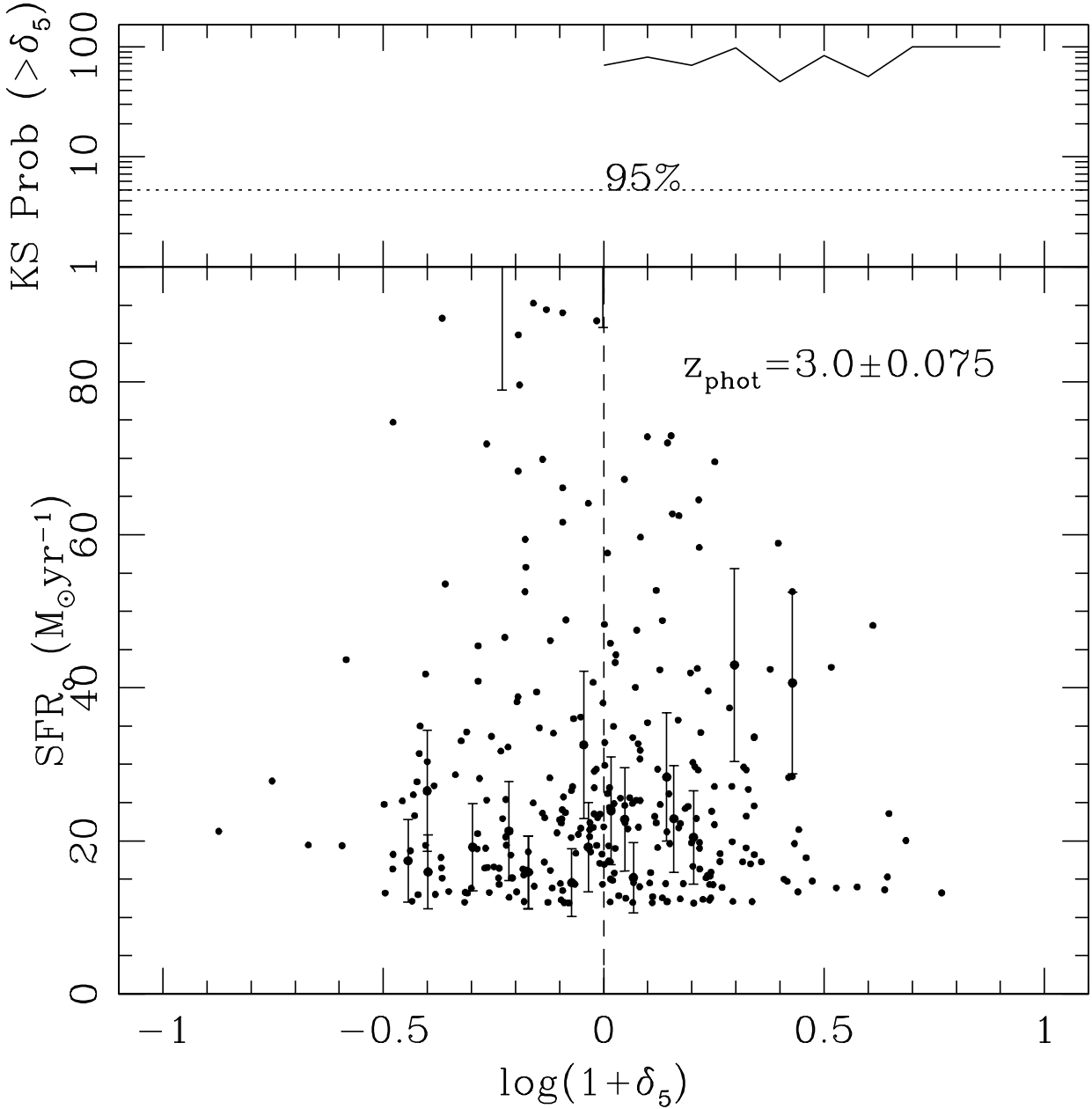}
\caption{{\it Bottom}: For LBGs selected to be at $z_{\rm phot}=3.0\pm0.075$ (see text),
we plot the dust-corrected SFR$_0$   against the
 over-density $\delta_5$ obtained from $\Sigma_5$, the surface density within the fifth nearest neighbor.
 Error bars on SFR are shown for 20 (randomly selected) LBGs.
 {\it Top}: Results for the KS-test. For a given over-density $\delta_5$, 
the probability that  the distribution of galaxies with over-densities $>\delta_5$ is drawn from the 
reference distribution (defined as $\delta_5<0$, {\it left of the vertical dashed line}) is shown.
The distribution in the high density bins is statistically the same as in the low density bins
(the 95\% rejection level is indicated by the dotted line).
 \label{fig:SFR-density} }
\end{figure}

Using the GIF2 simulations of section~\ref{section:simulations},
 we performed Monte-Carlo simulations to test whether or not the scatter in the $\Sigma_5$--$\rho_5$
 relation is responsible for our  null-result.
We selected randomly about a third of the simulated LBGs to match our sample size of \ngal\ LBGs.
These simulated LBGs were assigned a SFR using SFR$_{\rm sim}=(\rho_5/<\rho_5>)^s <{\rm SFR}>$, where
$\rho_5$ is the same as in Fig.~\ref{fig:compare:n5:n10}, $s$ is the slope of the simulated SFR-density relation,
 and $<{\rm SFR}>$ is the observed mean SFR in our data.
Noise with the same properties as in the data (i.e. $\sigma(\rm{SFR})\propto$SFR) was added to SFR$_{\rm sim}$.
Note that these mock catalogs have  the scatter between  $\Sigma_5$ and $\rho_5$ (Fig.~\ref{fig:compare:n5:n10}) built in.
We repeated our analysis of SFR$_{\rm sim}$ versus $\Sigma_5$ 1000 times for values of $s$ spanning $[-0.5,0.1]$.

 At $z=0$, the SFR decreases by a factor of 5 over 2 decade in density \citep{GomezP_03a,BaloghM_04a}, 
 i.e. $s\simeq \log(1/5)/2\simeq -0.35$.
In this case, the Pearson correlation test gives a probability smaller than $10^{-6}$  (or $\sim5$-$\sigma$)
of finding no correlation.
Even if we artificially increase the noise   by a factor as large as 2.5, as if we had underestimated our 
errors by such a factor, we find that there is less than a 0.1\%\ chance of finding no correlation if the $z=3$ SFR-density relation
were as steep as at $z=0$.
  If the $z=3$ SFR-density relation were  not as steep, but were 3.6 times flatter ($s=-0.07$),
   we would still have detected a correlation at the 2-$\sigma$ level.

\begin{figure}
\plottwo{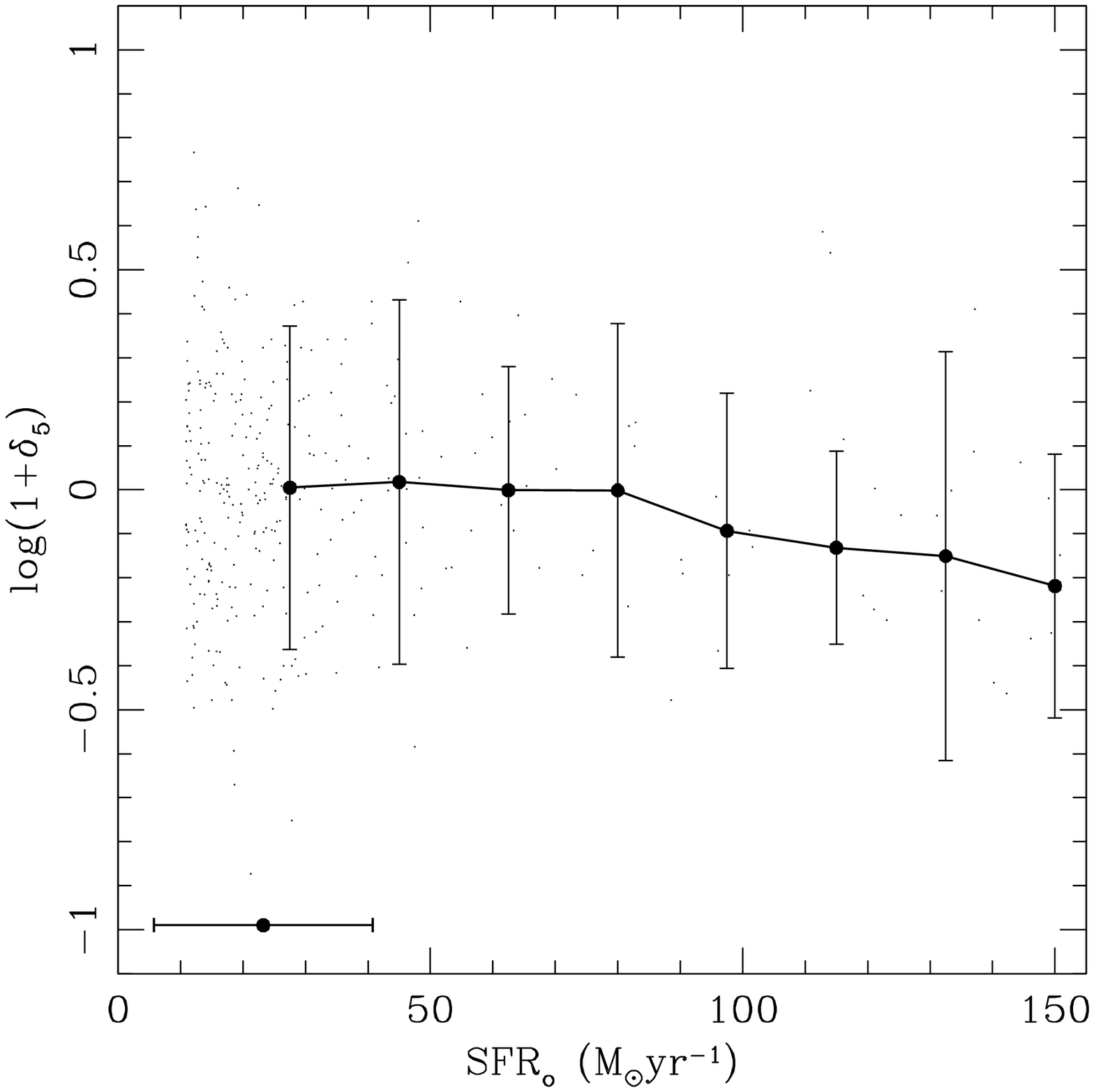}{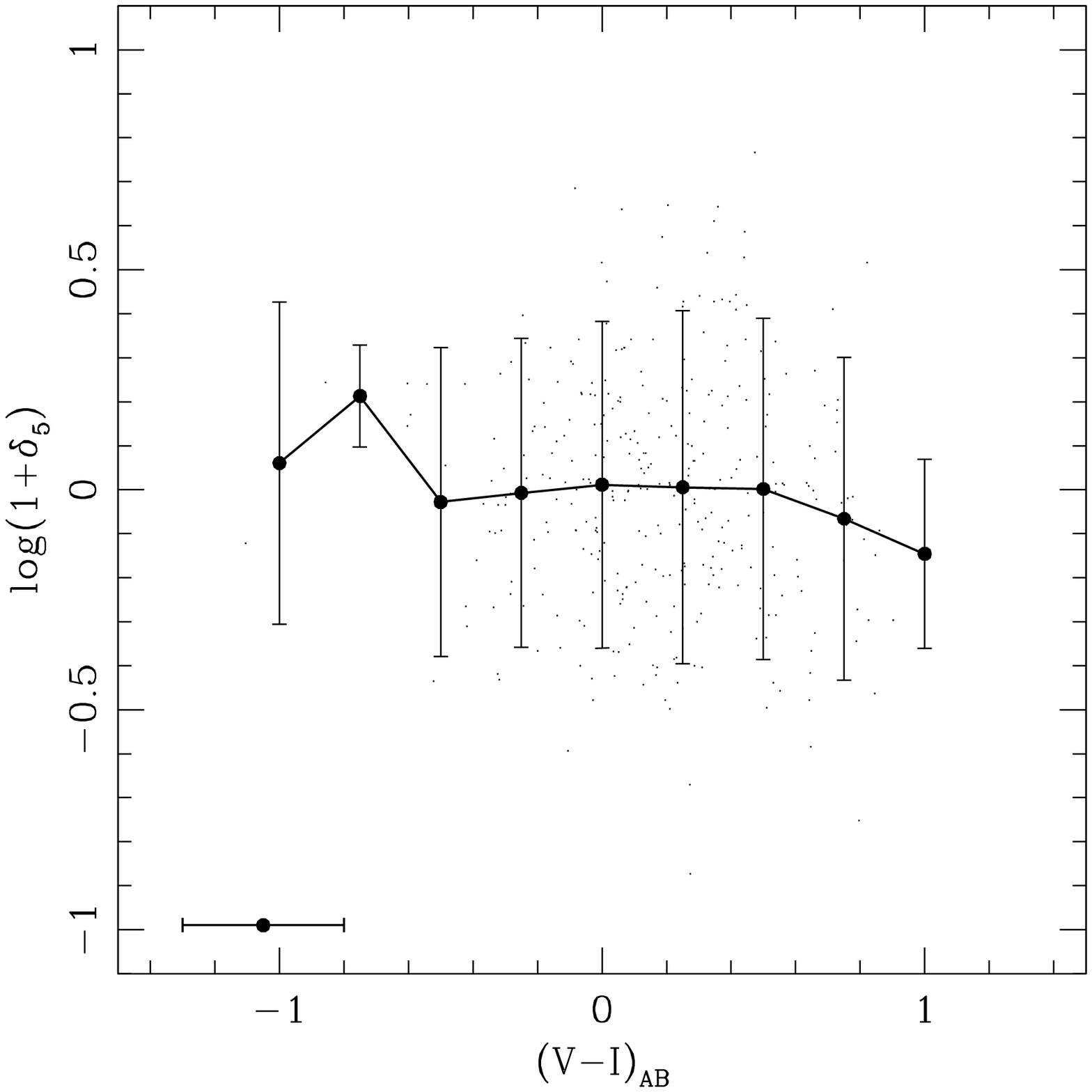}
\caption{{\it Left}: The  density-SFR relation at $z=3$. The filled circles show 
the median over-density $\delta_5$ in bins of  SFR$_0$ (bin size shown by the horizontal error bar).  
{\it Right}: Density-color relation at $z=3$. The filled circles show
the median over-density $\delta_5$ in color bins (bin size shown by the horizontal error bar).
 In both panels,
the thick  line  shows the running median, and the vertical error bars the 1-$\sigma$ spread, respectively.
The colors and SFR of LBGs are independent of the density of the surrounding galaxy.
 \label{fig:density-SFR} }
\end{figure}

As advocated   by \citet{HoggD_03a}, because the local density
has  a signal-to-noise ratio much lower than the other physical quantities,
it may be preferable to compute the mean
density at constant galaxy properties (e.g., SFR, $M$, color, etc.).
The left panel of Fig.~\ref{fig:density-SFR} shows the   over-density $\delta_5$ as a function of SFR.
The running median and the 1-$\sigma$ spread are shown by the thick line and the error bars, respectively.
Again, this plot reveals no evidence of an SFR-density relation.

We also investigate the dependence of galaxy color on density.
The right panel of Fig.~\ref{fig:density-SFR} shows the  over-density $\delta_5$ as a
function of the observed-frame $V-I$. Again, the running median and the 1-$\sigma$ spread are shown.
The rest-frame UV colors of LBGs do not appear to change with environment.
We note that  this non dependence of rest-frame UV color for blue galaxies such as LBGs
is similar to the results of $z=0$ surveys:
 in contrast to galaxies on the red sequence \citep[e.g.][]{BellE_04a},
the mean color of blue galaxies is
only weakly dependent on environment \citep{BlantonM_03b,HoggD_03a,HoggD_04a,BaloghM_04b}.
However, we cannot rule out an  environment dependence of the mean {\it rest-frame} optical colors (or age)
of  LBGs  given that we sample only $\lambda_{\rm rest}\la 2000$\AA.
Future rest-frame optical observations should
 address possible variations of the rest-frame optical colors of LBGs with environment.

\section{Summary and Conclusions}
\label{section:discussion}

From our wide-field images, covering a total of 0.90 deg$^2$, 
we selected  $z\simeq 3$ LBGs in several redshift
  slices  $100h^{-1}$~Mpc (co-moving) deep spanning 
    the redshift range  $2.9<z<3.4$.
 We computed the  SFR from the UV luminosity at rest wavelength $2000$~\AA.
 Using mock catalogs from the GIF2 simulations, we show that our photometric redshift accuracy  
  is sufficient to statistically distinguish between low- and high-density regions using
  projected density estimators such as the density within the fifth nearest neighbor $\Sigma_5$. 
 Our main results are as follows:  we find that  (1) there is no evidence of an SFR-density relation at $z=3$, and
 (2) the rest-frame UV colors of LBGs do not appear to change with environment.  
 
 Using Monte-Carlo simulations and the same projected density estimators that we applied to our data,
 we find  a probability  smaller than $10^{-6}$ of finding no correlation  
   if the steep $z=0$ SFR-density relation  were present in our data.
If the $z=3$  SFR-density relation were about four times flatter than the $z=0$  relation, 
we would still have detected a correlation at the 2-$\sigma$ level.
  We conclude that, unless the $z=3$ SFR-density relation is even flatter,
we would have detected it in our data.
  This is in sharp contrast to surveys at $z=0$ and $z\sim0.4$ 
      that have shown that the mean SFR is strongly dependent on
the  local galaxy density \citep[e.g.,][]{LewisI_02a,GomezP_03a,BaloghM_04a,KodamaT_04a}.

Very few predictions have been made of how the SFR-density relation should evolve from 
$z\simeq 3$ to the present. Recently,
\citet{KeresD_04a}, using smooth particle hydrodynamic cosmological simulations, showed that 
 the SFR-density relation is expected to be present as early as $z\sim2$, but not at $z\sim 3$ (see their Fig.~13).
See \citet{KeresD_04a} and \citet{BirnboimY_03a} for a detailed description of the 
physical mechanisms at play.
Our observed non-dependence of SFR on environment
supports the theoretical results of \citet{KeresD_04a}.

Naturally, our sample is not representative of the entire galaxy population at $z=3$;
it is biased towards blue star-forming galaxies and does not include
the red population unveiled by the Faint Infrared Extragalactic Survey \citep{LabbeI_03a}.
With the present data, we are unable to rule out the existence of a SFR-density relation
if one were to include this redder population.

\acknowledgments
We thank the anonymous referee for his/her comments that led to an
improved analysis.
We thank  D. Kere\v{s} for discussions, and
 D. Croton  for providing the catalog of the GIF2 simulation.
N.~B. acknowledges partial support from the European Community Research and Training Network 
 ``The Physics of the Intergalactic Medium.''
  J.~D.~L.  acknowledges support from NSF grant AST-0332504.



\end{document}